\documentclass[aps,prd,twocolumn,showpacs=false,superscriptaddress,nofootinbib]{revtex4-2}
\usepackage{amsmath,amssymb,mathtools,graphicx,bm,xcolor,hyperref}

\hypersetup{
setpagesize=false,
bookmarksnumbered=true,
bookmarksopen=true,%
colorlinks=true,%
linkcolor=blue,
citecolor=blue,
}

\newcommand{\N}{\mathcal N}
\newcommand{\NI}{\N_{A:B}}
\newcommand{\NE}{\N_{AB:CD}}
\newcommand{\Tr}{\operatorname{Tr}}

\begin{document}
\title{Complete characterization of internal–external entanglement trade-offs\\ in two-mode Gaussian states}
\author{Yasusada Nambu}
\email{yasusada.nambu@gmail.com}
\affiliation{%
Department of Physics, Graduate School of Science, Nagoya University, Nagoya
464-8602, Japan
}%
\affiliation{%
Kansai Institute for Photon Science (KPSI), National Institutes for Quantum Science and Technology (QST), Kizugawa,
Kyoto 619-0215, Japan
}%
\thanks{Visiting Collaborative Researcher at KPSI, QST}
\date{\today}
\begin{abstract}
We characterize the exact trade-off between internal and external entanglement in asymmetric two-mode Gaussian states at fixed local symplectic eigenvalues. Internal entanglement is quantified by the negativity $\N_{A:B}$ between the two modes, while external entanglement is quantified by the negativity $\N_{AB:CD}$ across a Gaussian purification. Numerical sampling suggests the boundary structure of the attainable region in the $(\N_{AB:CD},\N_{A:B})$ plane, which we then determine analytically. Parametrizing the global symplectic spectrum yields a closed admissible interval for the remaining spectral parameter at fixed $\N_{AB:CD}$. We prove constructively that every point in this interval corresponds to a physical covariance matrix, implying that all intermediate values of $\N_{A:B}$ between the resulting sharp bounds are achievable. The upper bound is attained for opposite-sign correlations of equal magnitude, while the positive branch of the lower bound occurs when one global symplectic eigenvalue equals unity. We also identify two external-entanglement thresholds, derive a saturated monogamy-type inequality, and recover the symmetric-state limit. Because the external negativity depends only on the spectrum of the reduced two-mode state, the bounds hold for any purification, including non-Gaussian purifications.
\end{abstract}
\maketitle

\section{Introduction}
Entanglement monogamy imposes constraints on how quantum entanglement can be distributed among the subsystems of a multipartite system. In its conventional formulation, the focus is on the sharing of entanglement across overlapping bipartitions. A representative example is the Coffman--Kundu--Wootters (CKW) inequality, which bounds the pairwise entanglement of a qubit with two other qubits by its entanglement with their joint system \cite{CKW}. Such inequalities illuminate the structure of multipartite entanglement, but their precise form and validity depend on the entanglement measure and the class of states considered \cite{DharReview}.

A conceptually distinct type of monogamy was developed by Camalet \cite{CamaletPRL2017,CamaletPRA2017,CamaletPRL2018}. Rather than comparing the entanglement shared by one subsystem with different partners, this approach examines the competition between quantum resources \emph{within} a composite subsystem and the correlations of that composite subsystem with its \emph{environment}. In particular, the external correlations in Camalet's general framework need not themselves be entanglement. The relevant partitions are therefore different from those in CKW-type entanglement-sharing inequalities: one resolves the internal structure of a selected composite system, while the other separates that system as a whole from its exterior. This distinction motivates the terminology \emph{internal--external correlation trade-off}.

Continuous-variable Gaussian states provide a tractable setting in which to quantify a specific internal--external entanglement trade-off. Their physicality and entanglement properties can be expressed in terms of covariance matrices and symplectic spectra \cite{Weedbrook,AdessoIlluminati,Adesso2014}. In this setting, Nambu and Yamaguchi investigated the relation between internal and external entanglement for symmetric two-mode Gaussian states, motivated by the emergence of classical behavior in quantum fields in de Sitter spacetime \cite{NY}. Unequal local symplectic eigenvalues, however, arise naturally when the two selected modes have different local mixedness, for example, due to unequal wave packets or detector responses.

Here we consider an asymmetric two-mode Gaussian state of modes $A$ and $B$, with fixed local symplectic eigenvalues $a$ and $b$. We quantify \emph{internal entanglement} by the negativity $\NI$ between $A$ and $B$, and \emph{external entanglement} by the negativity $\NE$ across the bipartition $AB:CD$ of a pure Gaussian purification. Unlike the general external correlations addressed in Camalet's framework, the latter quantity measures entanglement specifically; for a pure purification, it depends only on the reduced state of $AB$. We derive the exact attainable region in the $(\NE,\NI)$ plane, obtaining sharp upper and lower bounds at fixed $a,b$ and proving constructively that every intermediate value is physically realized. The upper bound gives a saturated internal--external entanglement trade-off, whereas the lower bound identifies regimes in which internal entanglement cannot vanish. Our results provide a complete characterization of this Gaussian-state trade-off, which is closely related to, but more specific than, Camalet's generalized monogamy framework. Although we use a two-mode Gaussian purifying system $CD$ for definiteness, our external negativity is independent of the purification: the same trade-off applies to any pure extension of the given two-mode Gaussian state. Throughout, $\hbar=1$ and logarithms are natural.

\section{Gaussian-state conventions and negativities}
For two bosonic modes, introduce canonical operators $\hat R=(\hat q_A,\hat p_A,\hat q_B,\hat p_B)^T$ with
\begin{equation}
[\hat R_j,\hat R_k]=i\Omega_{jk},\qquad \Omega=J\oplus J,\quad J=\begin{pmatrix}0&1\\-1&0\end{pmatrix}.
\end{equation}
We define the covariance matrix without a factor of $1/2$:
\begin{equation}
V_{jk}=\langle\{\hat R_j-\langle \hat R_j\rangle,\hat R_k-\langle \hat R_k\rangle\}\rangle.
\end{equation}
Thus, the uncertainty principle reads $V+i\Omega\geq0$, and the vacuum symplectic eigenvalue is unity. By local symplectic transformations, any two-mode Gaussian covariance matrix can be put into the standard form
\begin{equation}\label{std}
V=\begin{pmatrix}a&0&c_1&0\\0&a&0&c_2\\c_1&0&b&0\\0&c_2&0&b\end{pmatrix},\qquad a,b\geq1.
\end{equation}
Here $a,b$ are the local symplectic eigenvalues, and $c_1,c_2$ are real intermode correlations. We call $a=b$ symmetric and $a\ne b$ asymmetric.

For a density operator $\rho_{XY}$, define the negativity
\begin{equation}\label{negdef}
\N_{X:Y}=\frac{\|\rho_{XY}^{T_Y}\|_1-1}{2},\qquad
\|M\|_1=\Tr\sqrt{M^\dagger M},
\end{equation}
where $T_Y$ denotes partial transposition on $Y$. In particular, the trace norm of a Hermitian operator equals the sum of the absolute values of its eigenvalues. Since partial transposition preserves the trace, Eq.~\eqref{negdef} equals the sum of the absolute values of the negative eigenvalues. It should not be confused with logarithmic negativity, $E_\N=\ln(1+2\N)$.

The standard symplectic invariants are
\begin{align}
\Delta&=a^2+b^2+2c_1c_2,\label{delta0}\\
D&=\det V=(ab-c_1^2)(ab-c_2^2).\label{det0}
\end{align}
The ordinary symplectic eigenvalues are
\begin{equation}\label{nu}
\nu_\pm^2=\frac{\Delta\pm\sqrt{\Delta^2-4D}}{2},\qquad \nu_+\geq\nu_-\geq1.
\end{equation}
By Williamson's theorem, a positive-definite covariance matrix admits a symplectic diagonalization, and its symplectic eigenvalues characterize the associated Gaussian thermal normal modes \cite{Williamson,Weedbrook,AdessoIlluminati}. For a positive-definite $V$, the uncertainty principle is equivalent to $\nu_-\geq1$. In invariant form, one may impose $\Delta\geq2$ and $D\geq\Delta-1$, together with the positive definiteness of $V$; the latter requires $ab-c_j^2>0$ for $j=1,2$. Indeed, $D-\Delta+1=(\nu_+^2-1)(\nu_-^2-1)$, and $\Delta\geq2$ selects the physical branch.

Partial transposition corresponds to $c_2\mapsto-c_2$. With $\widetilde\Delta=a^2+b^2-2c_1c_2$, the internal negativity is
\begin{align}
\label{pt}
&\NI=\frac12\max\{\widetilde\nu_-^{-1}-1,0\},\\
&\widetilde\nu_-^2=\frac{\widetilde\Delta-\sqrt{\widetilde\Delta^2-4D}}2,
\end{align}
where $\widetilde\nu_{-}$ is the smaller partially transposed symplectic eigenvalue. 
For two-mode Gaussian states, the positive-partial-transpose (PPT) criterion is necessary and sufficient for separability \cite{Simon,Duan,AdessoIlluminati}.

Let $ABCD$ be a pure Gaussian purification of $AB$, with $CD$ as the purifying subsystem. The external negativity is
\begin{equation}
\label{external}
1+2\NE=\prod_{j=\pm}\left(\nu_j+\sqrt{\nu_j^2-1}\right).
\end{equation}
More generally, let $E$ be an arbitrary purifying system and $|\Psi\rangle_{ABE}$ any pure state with $\operatorname{Tr}_E|\Psi\rangle\langle\Psi|=\rho_{AB}$. Its Schmidt coefficients across $AB:E$ are the square roots of the eigenvalues of $\rho_{AB}$, and therefore
\begin{equation}\label{purification_invariance}
\N_{AB:E}=\frac12\left[\left(\Tr\sqrt{\rho_{AB}}\right)^2-1\right].
\end{equation}
Thus $\N_{AB:E}=\N_{AB:CD}$, irrespective of the dimension or Gaussian character of the purifying system. In particular, every attainable-region bound and threshold below applies to arbitrary purifications of a two-mode Gaussian $\rho_{AB}$.

A general two-mode Gaussian state with $\nu_+>1$ and $\nu_->1$ requires two ancillary bosonic modes for a \emph{Gaussian} purification; if precisely one symplectic eigenvalue exceeds unity, one Gaussian ancillary mode suffices, and if both equal unity no ancilla is needed. By contrast, a single bosonic mode has a countably infinite-dimensional Hilbert space and can purify any two-mode Gaussian state if a non-Gaussian joint pure state is allowed. These facts concern the realization of a purification, not the value of Eq.~\eqref{purification_invariance}.

For completeness, Appendix~\ref{app:external} derives this expression directly from the Schmidt spectrum, including the normalization of the trace norm.

\section{Physical regions and numerical evidence for entanglement bounds}\label{sec:physical_numerical}
\subsection{Physical and separable regions}
\begin{figure*}[t]
\includegraphics[width=.94\textwidth]{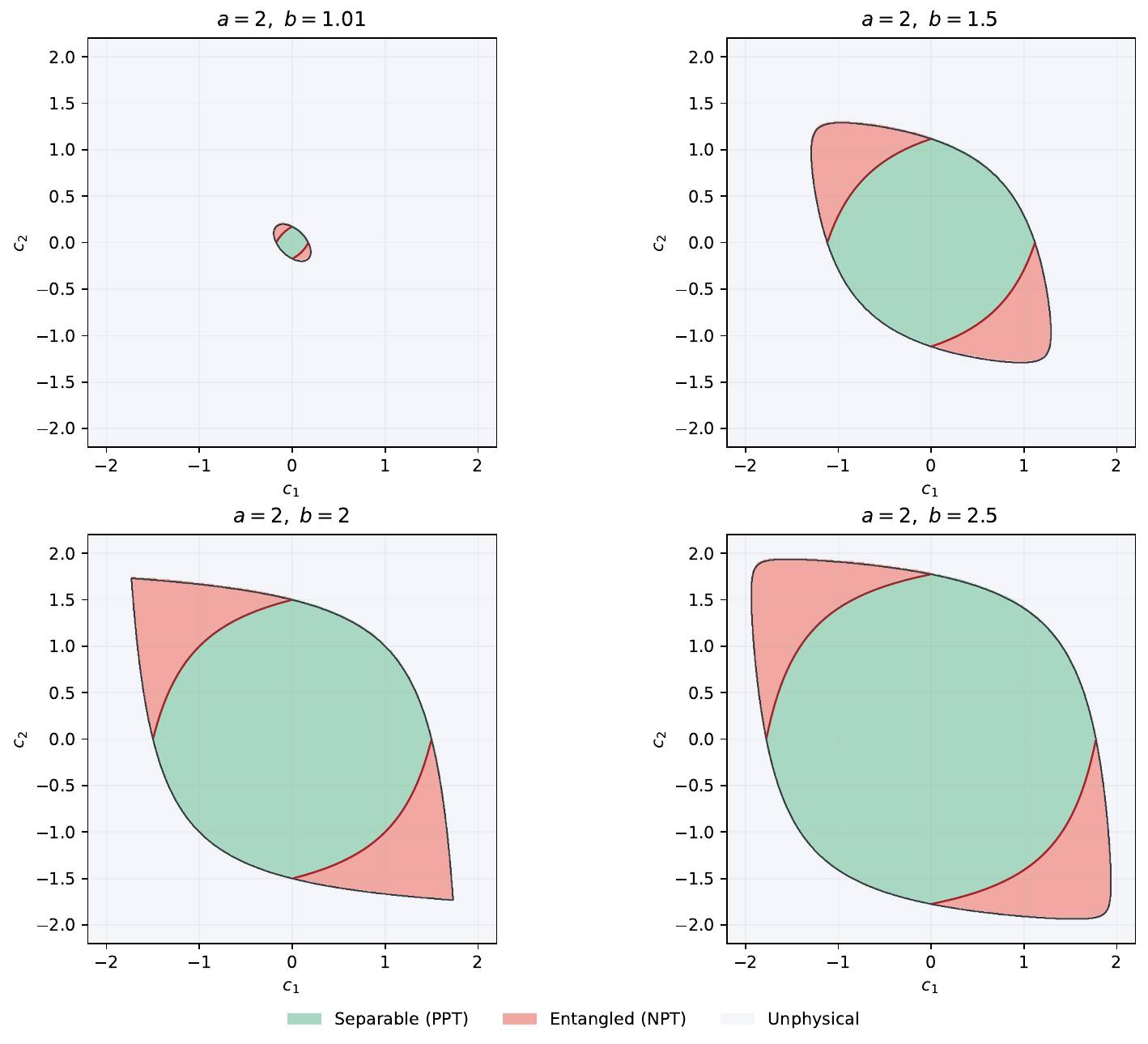}
\caption{Physicality and separability in the $(c_1,c_2)$ plane for $a=2$ and $b=1.01,1.5,2,2.5$. Green denotes physical PPT (separable) states, red denotes physical NPT (entangled) states, and pale gray denotes unphysical covariance matrices. The outer boundary is set by physicality, while the internal boundary is given by $\widetilde\nu_-=1$. The axes have the same range in all panels.}\label{fig:phys}
\end{figure*}

Figure~\ref{fig:phys} shows the physically allowed standard-form correlations $(c_1,c_2)$ for $a=2$ and several values of $b$. The PPT and non-PPT (NPT) regions are distinguished within the physical domain. As $b\to1$, the reduced state of mode $B$ approaches a pure Gaussian state. Consequently, the joint state factorizes as $\rho_{AB}=\rho_A\otimes\rho_B$, and the internal entanglement vanishes. A pure two-mode Gaussian state in the standard form exists only when $a=b$; its correlations are $c_1=-c_2=\pm\sqrt{a^2-1}$, corresponding to two-mode squeezed vacua up to local transformations.

\subsection{Numerical exploration and conjectured bounds}
\begin{figure*}[t]
\includegraphics[width=.93\textwidth]{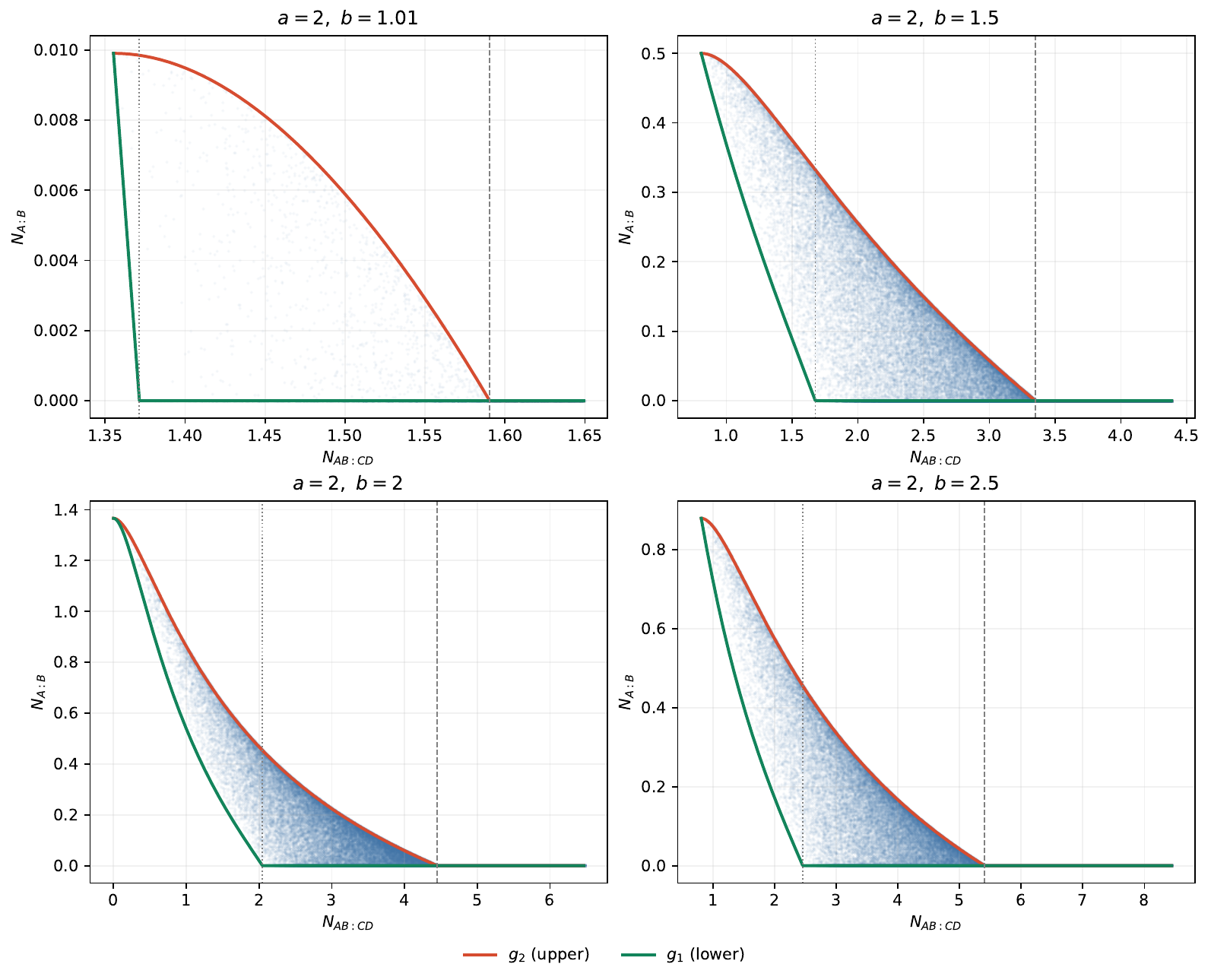}
\caption{Internal negativity $\NI$ versus external negativity $\NE$ for $a=2$ and $b=1.01,1.5,2,2.5$. Blue points are physical standard-form states drawn from $290\,000$ uniformly sampled pairs $(c_1,c_2)$ per panel before physicality selection; the respective numbers of retained points are $3733$, $114998$, $156064$, and $171204$. Red and green curves show the proposed upper and positive lower envelopes $g_2$ and $g_1$, whose exactness and attainability are established analytically below. The dotted and dashed vertical lines mark $\alpha$ and $\beta$, respectively.}\label{fig:scatter}
\end{figure*}

To complement the physicality and separability domains shown above, we examine the image of the physical covariance-matrix domain under the map $(c_1,c_2)\mapsto(\NE,\NI)$. For each choice of local symplectic eigenvalues, we uniformly sample $290\,000$ pairs $(c_1,c_2)$, retain only positive-definite matrices satisfying $V+i\Omega\geq0$, and evaluate both negativities using Eqs.~\eqref{pt} and \eqref{external}. Figure~\ref{fig:scatter} shows the resulting scatter plots. This numerical procedure provides evidence for the geometry of the attainable region, but cannot by itself prove that its apparent boundary is sharp or that the interior contains no gaps.

The numerical observations suggest a conjecture: at fixed $a,b$, the attainable pairs $(\NE,\NI)$ form a closed region between two continuous envelopes, without inaccessible gaps. Three features motivate the analytic investigation. First, the upper edge appears to follow a one-parameter family with $c_1=-c_2$, while the positive portion of the lower edge appears to be associated with the smaller global symplectic eigenvalue $\nu_-$ equal to one. Second, there are two distinct external-negativity thresholds: the lower edge first reaches zero, and at a larger external negativity the upper edge also vanishes. Third, the populated region appears to fill the space between the two edges. The overlaid curves are evaluated from the analytic expressions presented in later sections and serve as guides to the numerical data; their appearance in the figure is not used as an assumption in the proof. The following analysis establishes these observations exactly, including the attainability of every intermediate value.

\section{Optimization at fixed external negativity}
Set
\begin{align}
&s=a+b,\quad d=|a-b|,\\
&n=\NE,\quad \nu_\pm=\cosh r_\pm,
\end{align}
where $r_+\geq r_-\geq0$. Introduce
\begin{align}
&r_++r_-=\ln(2n+1),\\
&C=\cosh(r_++r_-),\quad z=\cosh(r_+-r_-).
\end{align}
Equation~\eqref{external} shows that fixing $n$ is equivalent to fixing $C$ because 
\begin{equation}
2n+1=e^{r_+}e^{r_-}=C+\sqrt{C^2-1}. 
\end{equation}
For fixed $a$ and $b$, our approach is to study how $\mathcal{N}_{A:B}$ and $\mathcal{N}_{AB:CD}$ depend on $(C,z)$, with $C\ge 1$ and $z\ge 1$. Standard hyperbolic identities yield
\begin{equation}
\label{CD}
\Delta=1+Cz,\qquad \sqrt D=\frac{C+z}{2}.
\end{equation}
Since $\widetilde\Delta=2(a^2+b^2)-\Delta$, the square of the smaller partially transposed symplectic eigenvalue is
\begin{align}
\label{lambda}
\lambda(z)&:=\widetilde\nu_-^2=\frac{T-\sqrt{T^2-(C+z)^2}}{2},\\
T&=2(a^2+b^2)-1-Cz.\nonumber
\end{align}
Keeping $C$ fixed, differentiating the defining quadratic equation $\lambda^2-T\lambda+(C+z)^2/4=0$ with respect to $z$ gives
\begin{equation}
\label{monotonic}
\frac{d\lambda}{dz}=\frac{C\lambda+(C+z)/2}{\sqrt{T^2-(C+z)^2}}>0.
\end{equation}
Degenerate cases follow by continuity. Thus, the internal negativity decreases monotonically with $z$ wherever it is nonzero. The problem reduces to finding the smallest and largest physically attainable $z$ at fixed $C$.

\subsection{Exact admissible interval and its attainability}\label{sec:attainability}
We next establish that the admissible values of $z$ form a full interval, a fact needed to show that the bounds on $\NI$ enclose no inaccessible gaps. We first derive the necessary spectral constraints rather than assuming them. For a positive-definite standard-form covariance matrix, $ab-c_j^2>0$, and the determinant in Eq.~\eqref{det0} obeys
\begin{equation}
\begin{split}
 &(ab-|c_1c_2|)^2-D\\
 &\quad\quad\quad =ab(|c_1|-|c_2|)^2\geq0.
\end{split}
\end{equation}
Because $ab-|c_1c_2|>0$, we obtain $\sqrt D\leq ab-|c_1c_2|$. Using $\Delta=a^2+b^2+2c_1c_2$ gives
\begin{align}
 (\nu_+-\nu_-)^2&=\Delta-2\sqrt D \notag\\
 &\geq (a-b)^2+2(c_1c_2+|c_1c_2|) \notag\\
 &\geq(a-b)^2,\\
 (\nu_++\nu_-)^2&=\Delta+2\sqrt D \notag\\
 &\leq(a+b)^2+2(c_1c_2-|c_1c_2|) \notag\\
 &\leq(a+b)^2.
\end{align}
Since all symplectic eigenvalues are nonnegative, these relations imply the following necessary conditions. The global and local symplectic spectra satisfy
\begin{equation}\label{marginalcompat}
\nu_+-\nu_-\geq d,\qquad \nu_++\nu_-\leq s,
\end{equation}
where $s=a+b$ and $d=|a-b|$. Together with $\nu_-\geq1$, these yield the range of $z$ with a fixed $C$:
\begin{equation}\label{zinterval}
 z_{\min}(C)\leq z\leq z_{\max}(C),
\end{equation}
with
\begin{align}
 z_{\min}(C)&=1+\frac{d^2}{C-1},\label{zminexact}\\
 z_{\max}(C)&=\min\left\{C,\frac{s^2}{C+1}-1\right\}.\label{zmaxexact}
\end{align}
Indeed, $(\nu_+-\nu_-)^2=(C-1)(z-1)$ and
$(\nu_++\nu_-)^2=(C+1)(z+1)$, while $\nu_-\geq1$ is equivalent to $z\leq C$. The endpoint $C=1$ is possible only when $a=b$, in which case the admissible interval collapses to $z=1$ and $\nu_+=\nu_-=1$. At this pure-state endpoint, Eq.~\eqref{zminexact} is replaced by $z_{\min}=1$; the apparent $0/0$ is not evaluated directly.

We now demonstrate sufficiency by explicitly reconstructing the standard-form covariance matrix~\eqref{std}. For prescribed $a$, $b$, $C$, and $z$, let $c_1$ and $c_2$ denote the real intermode correlation coefficients appearing in Eq.~\eqref{std}; in particular, they specify the position--position and momentum--momentum covariances, respectively. Introduce their product and sum of squares,
\begin{equation}
 u=c_1c_2,\qquad w=c_1^2+c_2^2.
\end{equation}
The symplectic invariants \eqref{delta0} and \eqref{det0}, together with Eq.~\eqref{CD}, determine these combinations as
\begin{align}
 u(z)&=\frac{1+Cz-a^2-b^2}{2},\label{uz}\\
 w(z)&=\frac{a^2b^2+u(z)^2-(C+z)^2/4}{ab}.\label{wz}
\end{align}
Solving $c_1c_2=u$ and $c_1^2+c_2^2=w$ then gives the explicit reconstruction
\begin{align}
 c_1^2&=\frac{w+\sqrt{w^2-4u^2}}{2},\label{constructcorr1}\\
 c_2^2&=\frac{w-\sqrt{w^2-4u^2}}{2}.\label{constructcorr2}
\end{align}
We choose the signs of $c_1$ and $c_2$ such that $c_1c_2=u$. Interchanging the two coefficients yields an equivalent choice of standard-form correlations. These expressions define real correlation coefficients provided that $w\geq2|u|$. We now show that this condition holds throughout the proposed interval and that the resulting covariance matrix satisfies the uncertainty principle.

The reality condition $w\geq2|u|$ is equivalent to the pair of inequalities $w+2u\geq0$ and $w-2u\geq0$. The spectral compatibility conditions imply $ab\pm u\geq(C+z)/2$, which in turn ensure both reality inequalities. To verify this explicitly, note that
\begin{equation}
 ab(w\pm2u)=(ab\pm u)^2-\frac{(C+z)^2}{4}.
\end{equation}
The inequalities
\begin{align}
 ab+u&\geq\frac{C+z}{2}
 \quad\Longleftrightarrow\quad (C-1)(z-1)\geq d^2,\label{reality1}\\
 ab-u&\geq\frac{C+z}{2}
 \quad\Longleftrightarrow\quad (C+1)(z+1)\leq s^2\label{reality2}
\end{align}
are precisely the two spectral compatibility conditions used to obtain Eq.~\eqref{zinterval}. They imply $w\geq2|u|$, so both reconstructed squares are real and nonnegative. Moreover, these conditions imply $|u|<ab$. The two factors $ab-c_j^2$ have a positive product, $D=(C+z)^2/4>0$, and their sum is also positive:
\begin{align}
 (ab-c_1^2)+(ab-c_2^2)&=2ab-w \notag\\
 &=\frac{a^2b^2-u^2+(C+z)^2/4}{ab}>0.
\end{align}
Here, the last inequality follows from $ab>|u|$ and $a,b\geq1$. Therefore, both factors are positive, $ab-c_j^2>0$ for $j=1,2$. The reconstructed $V$ is therefore positive definite. Finally, its symplectic invariants reproduce Eq.~\eqref{CD}, while $z\leq C$ guarantees $\nu_-\geq1$. Consequently, the uncertainty principle holds, and the reconstructed covariance matrix is physical. This establishes the attainability of every $z$ in Eq.~\eqref{zinterval} without assuming the connectedness of the original $(c_1,c_2)$ domain.

The interval in Eq.~\eqref{zinterval} is nonempty if and only if
$z_{\min}(C)\leq z_{\max}(C)$. For $C>1$, the first requirement,
$z_{\min}(C)\leq C$, gives
\begin{equation}
1+\frac{d^2}{C-1}\leq C
\quad\Longleftrightarrow\quad C\geq1+d.
\end{equation}
The second requirement, $z_{\min}(C)\leq s^2/(C+1)-1$, is
\begin{equation}
1+\frac{d^2}{C-1}\leq\frac{s^2}{C+1}-1.
\end{equation}
Multiplying by the positive quantity $(C-1)(C+1)$ and using
$s^2-d^2=4ab$, we find
\begin{equation}
C^2-2abC+(a^2+b^2-1)\leq0.
\end{equation}
The two roots of this quadratic polynomial are
\begin{equation}
C_{\pm}=ab\pm\sqrt{(a^2-1)(b^2-1)}.
\end{equation}
Thus $C_-\leq C\leq C_+$. For $a,b\geq1$, the lower root obeys
$C_-\leq1+|a-b|=1+d$: for example, assuming $a\geq b$,
\begin{equation}
\begin{split}
C_- -(1+a-b)
&=(b-1)(a+1)\\
&\quad-\sqrt{(a^2-1)(b^2-1)}\leq0,
\end{split}
\end{equation}
where the last inequality follows by squaring both nonnegative
terms and using $a\geq b$. Therefore, the stronger lower bound is
$C\geq1+d$, and the combined conditions yield the exact range of $C$:
\begin{equation}\label{Crangeexact}
1+d\leq C\leq C_{\max},\quad
C_{\max}=ab+\sqrt{(a^2-1)(b^2-1)},
\end{equation}
with the $C=1$ endpoint included when $d=0$. The upper endpoint of $z$ changes branch at $C=s-1$. The two competing upper bounds in Eq.~\eqref{zmaxexact} coincide when $C=s^2/(C+1)-1$, or equivalently $(C+1)^2=s^2$. Since $C+1>0$, the crossing occurs at $C=s-1$. Thus the active boundary is $z_{\max}=C$ for $C\leq s-1$, and $z_{\max}=s^2/(C+1)-1$ for $C\geq s-1$.

The first branch corresponds to $\nu_-=1$: indeed, $C-z=2\sinh r_+\sinh r_-$, and $r_+\geq r_-\geq0$ implies $z=C$ if and only if $r_-=0$. The second branch saturates $\nu_++\nu_-=s$ and is attained by the same-sign equal-correlation family $c_1=c_2$, including the product state at $C=C_{\max}$. This crossing changes the \emph{physical boundary at fixed $C$}; it is not an additional entanglement threshold.

Since $\NI$ is continuous and nonincreasing in $z$, the continuous image of the entire interval \eqref{zinterval} is exactly the interval between its endpoint values. Thus, every internal negativity between the lower and upper bounds is realized at fixed $a,b,n$. Having established the exact admissible interval, we can determine the sharp negativity bounds by evaluating $\NI$ at its two endpoints.

\subsection{Tight upper bound}\label{sec:upper_bound}
Because $\NI$ is nonincreasing in $z$ by Eq.~\eqref{monotonic}, its maximum at fixed $C$ is attained at the admissible minimum $z=z_{\min}(C)$. As established in Sec.~\ref{sec:attainability}, this endpoint is realized by the opposite-sign equal-correlation family $c_1=-c_2=c$. Along this branch,
\begin{align}
\nu_\pm&=\frac12\left(\sqrt{s^2-4c^2}\pm d\right),\\
\widetilde\nu_-&=\frac12\left(s-\sqrt{d^2+4c^2}\right),\label{vopt}\\
c^2(n)&=ab-\frac12\left(C+1+\frac{d^2}{C-1}\right).\label{copt}
\end{align}
Define $v_2(n)=[s-\sqrt{d^2+4c^2(n)}]/2$. Then, the exact upper envelope is
\begin{equation}
\label{g2}
g_2(n;a,b)=\frac12\max\{v_2(n)^{-1}-1,0\}.
\end{equation}
The expression at $a=b$, $n=0$ is understood by continuity.

\subsection{Tight lower bound}
At fixed $C$, Eq.~\eqref{zmaxexact} shows that the largest admissible $z$ is not always $C$: the active constraint changes at $C=s-1$, as explained above. Since the internal negativity is nonincreasing in $z$, its minimum is reached at $z=z_{\max}(C)$. For $C\leq s-1$, the active endpoint is $z_{\max}=C$, equivalent to $r_-=0$ and $\nu_-=1$. On this branch, $\nu_+=C$, $D=C^2$, and $\Delta=C^2+1$. We therefore define
\begin{align}
v_1(n)^2&=\frac{T_1-\sqrt{T_1^2-4C^2}}2,\label{v1} \\
T_1&=2(a^2+b^2)-1-C^2,
\end{align}
and the lower-envelope function on the $\nu_-=1$ branch is
\begin{equation}
g_1(n;a,b)=\frac12\max\{v_1(n)^{-1}-1,0\}.\label{g1}
\end{equation}
The admissible range of this branch follows from $\nu_+=C$, $\nu_-=1$ and the compatibility conditions \eqref{marginalcompat}:
\begin{equation}\label{rangeC}
1+d\leq C\leq s-1.
\end{equation}
Its attainability follows from the construction in Sec.~\ref{sec:attainability} at $z=C$. For $C>s-1$, setting $z=C$ would imply $\nu_++\nu_-=C+1>s$, which is forbidden. The minimizing endpoint then saturates $\nu_++\nu_-=s$ and is realized by $c_1=c_2$. As established in the next section, the lower negativity has already vanished at $C=C_\alpha\leq s-1$. Thus the change of branch does not introduce a new entanglement threshold.

\section{Complete attainable region and entanglement thresholds}
We now determine the full range of external negativity on the upper-envelope family $c_1=-c_2=c$, corresponding to $z=z_{\min}(C)$. Along this family,
\begin{equation}
\nu_-=\frac{\sqrt{s^2-4c^2}-d}{2}.
\end{equation}
Consequently, the uncertainty principle gives
\begin{align}
\nu_-\geq1&\ \Longleftrightarrow\ \sqrt{s^2-4c^2}\geq d+2\\
&\ \Longleftrightarrow\ c^2\leq\frac{s^2-(d+2)^2}{4}\\
&\ \Longleftrightarrow\ c^2\leq ab-1-d.
\end{align}
Together with $c^2\geq0$, this yields
\begin{equation}
0\leq c^2\leq ab-1-d.
\end{equation}
Along this maximizing family, both symplectic eigenvalues $\nu_\pm$ decrease with increasing $c^2$ [Sec.~\ref{sec:upper_bound}], while $\widetilde\nu_-$ also decreases [Eq.~\eqref{vopt}]. Thus the external negativity decreases and the internal negativity increases with $c^2$, establishing that the upper envelope $g_2(n)$ is nonincreasing in $n$. The full allowed interval of $n$ is
\begin{align}
&n_{\min}=\frac12\left[d+\sqrt{d(d+2)}\right],\label{nmin}\\
&n_{\max}=\frac12\left[(a+\sqrt{a^2-1})(b+\sqrt{b^2-1})-1\right].
\label{nmax}
\end{align}
At $n_{\min}$, $\nu_-=1$ and $\nu_+=1+d$; at $n_{\max}$, $c_1=c_2=0$. Equivalently, $n_{\max}=[C_{\max}+\sqrt{C_{\max}^2-1}-1]/2$, since $C=\cosh(\operatorname{arcosh}\nu_++\operatorname{arcosh}\nu_-)$. Every intermediate $n$ is realized on the $c_1=-c_2$ branch, establishing global sharpness of $g_2$.

Two distinct thresholds characterize the entanglement structure within this interval: at $n=\alpha$ the minimum internal negativity first reaches zero, whereas at $n=\beta$ even its maximum becomes zero. We derive them in turn and establish the ordering $n_{\min}\leq\alpha\leq\beta\leq n_{\max}$.

The positive lower bound disappears when $v_1=1$. Substituting this into Eq.~\eqref{v1} gives $C_\alpha=\sqrt{a^2+b^2-1}$, or
\begin{equation}
\label{alpha}
\alpha(a,b)=\frac12\left[\sqrt{a^2+b^2-1}+\sqrt{a^2+b^2-2}-1\right].
\end{equation}
This threshold satisfies $g_1(\alpha)=0$. Moreover,
\begin{equation}
(s-1)^2-C_\alpha^2=2(a-1)(b-1)\geq0,
\end{equation}
so $C_\alpha\leq s-1$: the lower negativity vanishes \emph{before} the active upper-$z$ boundary changes at $C=s-1$ (with coincidence when $a=1$ or $b=1$). For $C_\alpha\leq C\leq s-1$, the $\nu_-=1$ endpoint itself is separable. For $C>s-1$, that endpoint ceases to be physical, but the new endpoint $\nu_++\nu_-=s$ is also separable: it can be realized with $c_1=c_2$, and such a standard-form covariance matrix is PPT. Consequently, the exact lower bound remains $\NI=0$ for every $n\geq\alpha$.

The upper bound vanishes at $\widetilde\nu_-=1$ on the branch $c_1=-c_2$, where
\begin{equation}
c_{\rm crit}^2=(a-1)(b-1).
\end{equation}
Define
\begin{equation}
w_\pm=\frac12\left[\sqrt{d^2+4(s-1)}\pm d\right].
\end{equation}
The second threshold is
\begin{equation}
\label{beta}
\beta(a,b)=\frac12\left[e^{\operatorname{arcosh}w_++\operatorname{arcosh}w_-}-1\right],
\end{equation}
and satisfies $g_2(\beta)=0$. For $a,b>1$, the ordering $\alpha<\beta$ follows directly from the admissible interval in Sec.~\ref{sec:attainability}: since
\begin{equation}
 C_\alpha^2-(1+d)^2
 =2\bigl(\min\{a,b\}-1\bigr)\bigl(\max\{a,b\}+1\bigr)>0,
\end{equation}
and $C_\alpha\leq s-1$, we have $z_{\min}(C_\alpha)<z_{\max}(C_\alpha)=C_\alpha$. At the upper endpoint $\widetilde\nu_-=1$, whereas strict monotonicity in $z$ [Eq.~\eqref{monotonic}] implies $\widetilde\nu_-<1$ at the lower endpoint. Hence $g_2(\alpha)>0$, and the upper envelope reaches zero only at the larger threshold $\beta$. For $a=1$ or $b=1$, the thresholds coincide.

The resulting sharp bounds describe the complete attainable region, not merely its envelope: the constructive proof in Sec.~\ref{sec:attainability} guarantees that every value between the two bounds is realized. They are
\begin{equation}
\label{piecewise}
\begin{cases}
g_1(n)\leq\NI\leq g_2(n),&n_{\min}\leq n\leq\alpha,\\
0\leq\NI\leq g_2(n),&\alpha\leq n\leq\beta,\\
\NI=0,&\beta\leq n\leq n_{\max}.
\end{cases}
\end{equation}
If $a=1$ or $b=1$, the state is a product state and all four thresholds coincide.

\section{Tight monogamy inequality and symmetric limit}
The largest possible internal negativity for fixed $a,b$ is attained on the $c_1=-c_2$ branch at $c^2=ab-1-d$:
\begin{equation}
\label{Nmax}
\N_{\max}(a,b)=\frac12\left[\frac{2}{s-\sqrt{s^2-4(1+d)}}-1\right].
\end{equation}
The tight upper bound can equivalently be written as a monogamy-type inequality. Define $\widetilde g(n;a,b)=\N_{\max}(a,b)-g_2(n;a,b)$. Since $g_2$ is nonincreasing in $n$, $\widetilde g$ is nonnegative and nondecreasing over the admissible interval. Then
\begin{equation}\label{monogamy}
\NI+\widetilde g(\NE;a,b)\leq\N_{\max}(a,b).
\end{equation}
Equation~\eqref{monogamy} provides a tight internal--external entanglement trade-off at fixed local symplectic eigenvalues. The family $c_1=-c_2$ saturates the inequality throughout the admissible range. For $n\geq\beta$, however, the upper bound $g_2(n;a,b)$ vanishes, so every physical state with the specified $a$, $b$, and $n$ saturates the inequality. Its structure is conceptually related to Camalet's generalized monogamy framework~\cite{CamaletPRL2017,CamaletPRA2017,CamaletPRL2018}, but it should not be confused with conventional CKW-type entanglement-sharing inequalities. In particular, it compares entanglement within the composite subsystem $AB$ with entanglement across the cut $AB:CD$, rather than two overlapping pairwise entanglements. Moreover, Eq.~\eqref{piecewise} gives the complete attainable region and reveals a regime $n<\alpha$ in which internal entanglement is unavoidable at the specified local mixedness and external negativity.

For $a=b$, $d=0$, $n_{\min}=0$, and
\begin{equation}
\N_{\max}(a,a)=\frac12\left(a+\sqrt{a^2-1}-1\right),
\end{equation}
recovering the symmetric-state trade-off studied in Ref.~\cite{NY}. In that case, the upper boundary is associated with $c_1=-c_2$, while the nonzero lower boundary lies on $\nu_-=1$.

\section{Discussion and conclusions}
Our results completely characterize the image of the physical two-mode Gaussian state space under the map $(c_1,c_2)\mapsto(\NE,\NI)$ at fixed $a,b$. Its central ingredient is the monotonicity \eqref{monotonic}, which turns the optimization into a comparison of attainable symplectic spectra. The numerical exploration in Fig.~\ref{fig:scatter} motivates the proposed boundary structure; the analytic construction proves its sharpness and the absence of inaccessible gaps.

An immediate application is to unequal localized field modes in curved spacetime, where the two local covariance blocks generally have different symplectic eigenvalues. Once the covariance matrix is reconstructed from two-point functions, Eqs.~\eqref{g2}--\eqref{piecewise} provide model-independent constraints on the distribution of entanglement between the observed pair and its purifying degrees of freedom. Applying these constraints to specific de Sitter, Unruh, or Hawking wave packets requires a separate field-theoretic calculation. A related application of entanglement trade-offs concerns the enhancement of correlations between localized field modes in moving-mirror radiation during intervals of negative energy flux \cite{Nambu2026}, which is related to information recovery in black hole evaporation. Establishing a quantitative connection with the present asymmetric bounds would require evaluating the corresponding local covariance matrices. Extensions to multimode and non-Gaussian states also require additional analysis since covariance data alone do not generally determine their negativity.

The two entanglement thresholds identify where the positive lower bound and then the upper bound disappear. More fundamentally, our results establish not only sharp bounds for arbitrary local asymmetry but also their complete attainability: every point between the bounds is realized by a physical Gaussian covariance matrix. This characterization goes beyond a monogamy inequality that merely excludes forbidden combinations of correlations. The use of two ancillary modes in a Gaussian purification is only a convenient representation: the external negativity is a spectral functional of $\rho_{AB}$ and is unchanged under arbitrary purifications. Consequently, the attainable region and its sharp boundaries do not depend on the number of purifying modes or on whether the purification is Gaussian. This invariance does not extend the present covariance-based optimization to non-Gaussian reduced states $\rho_{AB}$.

\appendix
\section{Derivation of the external negativity}\label{app:external}
Consider a pure bipartite state with Schmidt decomposition
\begin{equation}
|\Psi\rangle=\sum_k\sqrt{p_k}\,|k\rangle_{AB}|k\rangle_{CD},\qquad \sum_kp_k=1.
\end{equation}
The partial transpose of $|\Psi\rangle\langle\Psi|$ has eigenvalues $p_k$ and, for each $k<l$, the pair $\pm\sqrt{p_kp_l}$. Hence
\begin{equation}
\label{puretrace}
\bigl\|(|\Psi\rangle\langle\Psi|)^{T_{CD}}\bigr\|_1
=\sum_kp_k+2\sum_{k<l}\sqrt{p_kp_l}
=\left(\Tr\sqrt{\rho_{AB}}\right)^2.
\end{equation}
This identity is independent of Gaussianity and of the chosen purification. The label $CD$ can therefore be replaced by an arbitrary purifying system $E$ without changing the expression.

By Williamson's theorem \cite{Williamson,Weedbrook}, the reduced Gaussian state $\rho_{AB}$ is unitarily equivalent to a product of thermal states with symplectic eigenvalues $\nu_+$ and $\nu_-$. For a single thermal mode with mean occupation $\bar n=(\nu-1)/2$, its eigenvalues are
\begin{equation}
p_m=(1-\xi)\xi^m,\qquad \xi=\frac{\nu-1}{\nu+1}.
\end{equation}
The geometric series gives
\begin{align}
\left(\sum_{m=0}^\infty\sqrt{p_m}\right)^2
&=\frac{1+\sqrt\xi}{1-\sqrt\xi} \notag \\
&=\nu+\sqrt{\nu^2-1}.
\end{align}
The Schmidt weights for the two-mode thermal product factorize. Applying Eq.~\eqref{puretrace} to both factors immediately yields Eq.~\eqref{external}. In particular, this proof does not require $a=b$ or a particular choice of Gaussian partner modes.


\end{document}